# Ultraconserved Sequences in the Honeybee Genome – Are GC-rich Regions Preferred?

Manoj Pratim Samanta

Systemix Institute, Los Altos, CA 94024

**Among all insect genomes, honeybee displays one of the most unusual patterns with interspersed long AT and GC-rich segments. Nearly 75% of the protein-coding genes are located in the AT-rich segments of the genome, but the biological significance of the GC-rich regions is not well understood. Based on an observation that the bee miRNAs, actins and tubulins are located in the GC-rich segments, this work investigated whether other highly conserved genomic regions show similar preferences. Sequences ultraconserved between the genomes of honeybee and *Nasonia,* another hymenopteran insect, were determined. They showed strong preferences towards locating in the GC-rich regions of the bee genome.**

The genome of honeybee *Apis mellifera* was recently sequenced to understand the molecular origin of insect eusociality [THGSC-06]. In addition to the observation on expansion of bee-related protein families, nucleotide level analysis of the genome revealed several puzzling features. The genome was highly AT-rich [THGSC-06] and showed larger GC variation than all other eukaryotic genomes analyzed by us [Samanta-07a]. Protein-coding genes preferred to locate in the AT-rich regions of the genome [THGSC-06, Elsik-07, Jorgensen-06]. Visual inspection of introns and third codons of selected genes showed that almost all bases were indiscriminately converted to A/T, and any unconverted G/C base possible survived due to selection pressure [Samanta-07b]. Those features in honeybee were not merely a consequence of its general AT-richness, because equally AT-rich *Tribolium* genome showed different internal features [Samanta-07a]. Surprisingly, even though some regions of the genome underwent such extreme conversion to A/T bases, other long GC-rich segments, covering nearly half of the bee genome, survived those changes. Biological significance of the GC-rich regions remains unclear.

It was observed that the miRNAs, actins and tubulins were located in the GC-rich regions of the bee genome [Weaver-07, Samanta-07c]. MiRNAs are short RNAs, whose sequences remain conserved between distant eukaryotic genomes. Similar nucleotide level conservation of segments of actin and tubulin genes were observed from sequence comparison between bee, fly, sea urchin and mouse genomes [Samanta-07c]. Therefore, this work investigated whether the ultraconserved sequences in bee, in general, preferred to locate in the GC-rich segments of the genome. The analysis was aided by recent sequencing of another hymenopteran insect, *Nasonia vitripennis*. Some of the internal characteristics of *Nasonia* and bee genomes are similar, suggesting that the unusual features in the bee genome were possibly present in their common ancestor [Samanta-07a]. However, their evolutionary distance is significantly large so that the neutral bases of the protein-coding genes do not remain generally conserved. This was confirmed by comparing third codon nucleotide levels among the bee and *Nasonia* genes that were highly conserved between the two insects [Data not shown].

Following a computational procedure (Methods), 714 nucleotide sequences, longer than 60 bases and ultraconserved between the bee and *Nasonia* genomes, were determined (Supplementary Table S1 available from http://www.systemix.org/reports/2/TableS1.txt). The sequences are 61-712 nucleotides long. For every sequence, additional 500 bases of flanking regions were included on each side to compute the GC level. Median GC level of the protein-coding genes, computed in similar manner, was 29% and was lower than the overall GC level of the bee genome (32%). In comparison, the median GC level of the ultraconserved regions (40%) was significantly higher than the overall genome. Fig. 1 compares the GC distributions of the protein-coding and ultraconserved regions. Locational biases of these two classes are clear from the picture. Among 714 ultraconserved regions, 116 matched exons of high confidence protein-coding genes in honeybee (Official or GLEAN set in Ref. [THGSC-06]). GC levels of those exons, determined in the same manner as above, was 37%, significantly higher than all protein-coding genes.

The above dichotomy between the protein-coding and ultraconserved regions of bee genome is very puzzling. Its significance regarding evolution of the bee genome can only be understood in the context of the bigger picture. Therefore, we first discuss the outstanding questions and evidences collected so far by different researchers, and then propose a hypothesis about evolution of the bee genome consistent with the presented evidences.

Following questions need to be answered.

1) How did the bee genome evolve to display such unique AT- and GC-rich regions [THGSC-06]? Jorgensen et al. discussed two alternatives [Jorgensen-06] – (1A) Entire honeybee genome experieneced strong mutational bias towards A/T nucleotides and any GC-regions was maintained through selection, or (1B) Different regions of the bee genome experieneced two distinct types of mutational patterns.

2) Did the common ancestor of bee and *Nasonia* evolve into such unique bimodal distribution (2A), or did the uniqueness develop in bee after differentiation of bee and *Nasonia* (2B) ?

Available evidences are as follows.

1. GC variations in both bee and *Nasonia* are higher than all other eukaryotic genomes. Also, the nucleotide distributions in both genomes show similar broad patterns, and the pattern was unlike any other eukaryotic organism studied by us [Samanta-07a]. This supports 2A.

2. *Nasonia* genome is as GC-rich as *Drosophila* or *Anopheles,* and not AT-rich like honeybee. Although this is apparently more supportive of 2B, we also observed large variation in GC level among different Dipteran insects. Therefore, it is possible that the common ancestor of bee and *Nasonia* was more bee-like, and then the overall GC level of *Nasonia* genome increased without modifying its GC-variation discussed in 1. The above explanation is satisfying except for one point. The third codons of *Nasonia* genes have higher GC than the overall genome, whereas the relationship is opposite in honeybee. Honeybee genome is the only eukaryotic genome analyzed by us that displays such opposite relationship between third codon nucleotide distribution and overall GC distribution [Samanta-07a]. Therefore, if 2A has to hold, whichever process led to the increase in overall GC level in *Nasonia* must have acted stronger on the protein-coding genes to increase the GC levels of their third codons even further.

3. Jorgensen et al. observed that the protein-coding genes from AT- and GC-rich regions in honeybee genome show different codon biases. This and a set of other observations regarding the locations and nucleotide-contents of protein-coding genes led them to conclude that 1B is valid [Jorgensen-06].

4. Neutral bases of protein-coding genes from AT-rich regions of the bee genome are extremely AT-rich. Introns and third-codons of some genes appear to be almost completely converted to A/T bases. This, in association with the fact, that nearly 75% of the bee genes are located in the AT-rich regions tends to support 1A, but leaves open the question about why some protein-coding genes still remain in the GC-rich segments.

5. Sequences ultraconserved between bee and *Nasonia* are more likely to locate in the GC-rich regions of the bee genome. This supports 1A, but still leaves open the question about why the relatively unconserved bases around the ultraconserved regions also remain GC-rich. For example, the stem-loops of the miRNAs and the unconserved third codons of actins and tubulins were free to convert to A/T under a global A/T conversion, but they did not.

6. Jorgensen *et al.* observed that the AT-rich regions are evolving more slowly than the GC-rich regions in honeybee. This was derived from an analysis of the protein-coding genes, and therefore may not reflect on the noncoding conserved regions.

Regarding 2A and 2B, most evidences are in stronger support of 2A. Regarding 1A and 1B, making a case for 1A leaves some questions unanswered. A case for 1B was made in Jorgensen *et al.*, except that the mechanism for why certain regions of the genome were preferred for AT-conversion than others was not clear. Based on the evidences presented here, we make the following hypothesis that merges 1A and 1B. The entire bee genome is under strong mutational bias towards A/T bases, but if a number of consecutive nucleotides in a region must stay GC-rich due to selection pressure, they become catalysts to convert a larger neighboring region to higher GC-level. This constraint is not present for protein-coding genes, where every third codon is mutable to A/T. However, if some protein-coding genes have conserved GC-rich third codons, or GC-dominated codons, their neighborhoods remain GC-rich. The above hypothesis can explain evidences 3-6 best, although the exact biochemical mechanism for such behavior is unclear at this moment.

In conclusion, the data on ultraconserved regions presented here suggest that the

GC-rich regions of the bee genome are neither empty, nor insignificant. They contain some key coding and noncoding genes that are under strong selection pressure. Because the *Nasonia* genome is more amenable to genetic manipulations than honeybee, genetic analysis of those regions in *Nasonia* may shed further light on their roles in hymenopteran biology.

## Methods

Twenty mer sequences were obtained by splitting both strands of the entire *Apis mellifera* and *Nasonia vitripennis* genomes, incremented by single bases. To avoid repetitive sequences, any 20 mer present more than 5 times in the combined set was discarded. From the remaining set, any 20 mer present in both genome was extracted into a new table. This final table contained all regions of length 20 base or more that were conserved between the two hymenopteran insects. A comparison between the V0.5 release of *Nasonia* genome and V2 release of *Apis mellifera* genome identified 10,117,610 instances of 20mers present in both genomes.

Once the above set is determined, it can be processed in different ways to identify longer conserved regions. This work used the following procedure. The set of conserved 20 mers was splitted among each scaffold pair from bee and *Nasonia*. They were sorted according to their genomic coordinates, and then clustered allowing a maximum of 5 consecutive gaps or mismatches in each cluster. From the clusters, all conserved regions longer than 60 nucleotides in the bee genome were collected. We note that slight modifications in procedure and parameters did not change the overall conclusion.

## Acknowledgments

This work is dedicated to Prof. Ryszard Maleszka of ANU for providing encouragement and helpful comments.

# Figures

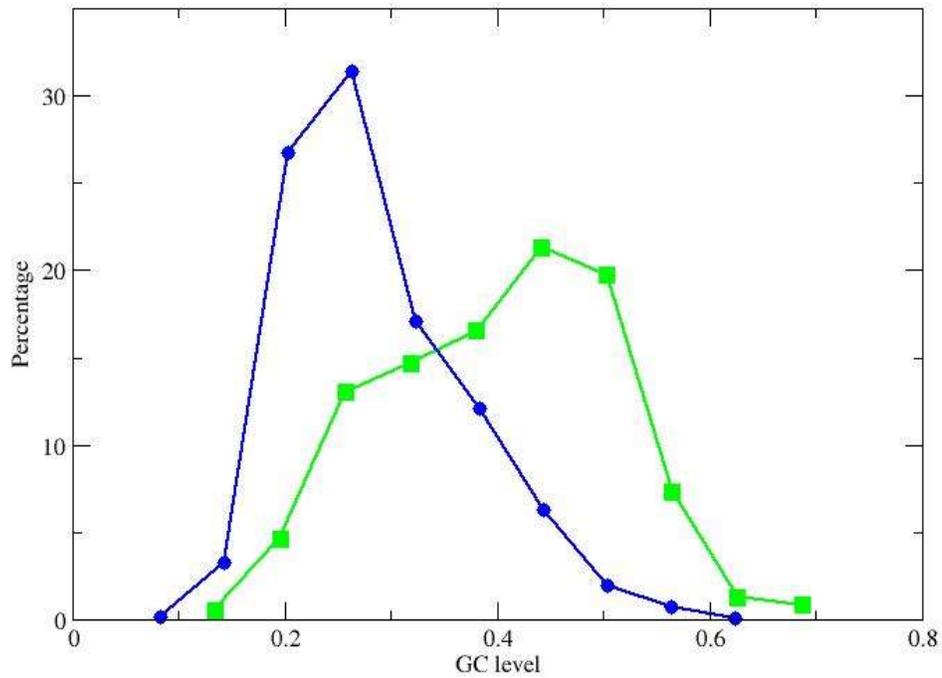

**Figure 1. Distribution of conserved regions.** GC-levels in bee genome for regions containing ultraconserved sequences (green) and those containing protein-coding genes (blue). Unlike protein-coding genes, ultraconserved regions are more likely to be located in the GC-rich segments of the bee genome. Overall GC level of the entire bee genome is ~32%.